# Affordances of Sketched Notations for Multimodal UI Design and Development Tools

Sam H. Ross, Yunseo Lee, Coco K. Lee, Jayne Everson, R. Benjamin Shapiro
*University of Washington*, Seattle, USA
{samhross, yunseol1, keliyk, everjay, rbs}@cs.washington.edu

*Abstract*—Multimodal UI design and development tools that interpret sketches or natural language descriptions of UIs inherently have notations: the inputs they can understand. In AI-based systems, notations are implicitly defined by the data used to train these systems. In order to create usable and intuitive notations for interactive design systems, we must regard, design, and evaluate these training datasets as notation specifications. To better understand the design space of notational possibilities for future design tools, we use the Cognitive Dimensions of Notations framework to analyze two possible notations for UI sketching. The first notation is the sketching rules for an existing UI sketch dataset, and the second notation is the set of sketches generated by participants in this study, where individuals sketched UIs without imposed representational rules. We imagine two systems, **FixedSketch** and **FlexiSketch**, built with each notation respectively, in order to understand the differential affordances of, and potential design requirements for, systems. We find that participants' sketches were composed of element-level notations that are ambiguous in isolation but are interpretable in context within whole designs. For many cognitive dimensions, the **FlexiSketch** notation supports greater intuitive creative expression and affords lower cognitive effort than the **FixedSketch** notation, but cannot be supported with prevailing, element-based approaches to UI sketch recognition. We argue that for future multimodal design tools to be truly human-centered, they must adopt contemporary AI methods, including transformer-based and human-in-the-loop, reinforcement learning techniques to understand users' context-rich expressive notations and corrections.

## I. Introduction

Multimodal tools for User Interface (UI) design and development have the potential to revolutionize how we, *and who can*, build software. At this moment in time, programming skills are a significant barrier to entry for software development. This doesn't have to be the case. Recent advances in machine learning have opened up the space of what is possible, enabling the creation of multimodal software design tools that interpret visual sketches and verbal or written descriptions of interfaces, instead of code, to generate software.

Though motivated by the long term question of *How can we create new multimodal design and development tools that provide creative and expressive computational agency?*, we have begun our empirical research on sketch-based input only. Sketch is a particularly relevant starting point for inquiry, as it is often the first phase of a design process— it is a low-cost way to express an idea without any of the constraints enforced by design tools or friction inherent in engineering. UI design professionals and amateurs alike often begin with sketches as part of initial ideation [1]–[4]. There are existing datasets of UI sketches and tools that aim to build functional UIs from these sketches. For these systems, sketch replaces programming languages or pixel-perfect mock-ups (eg. Figma) as the notation of design. In the context of ML-based sketch recognition systems, a training dataset of UI sketches is inherently a notation, even if a model's learned representations are not explainable. These systems will only correctly interpret sketches like those they are trained on, meaning a user has to use these representations to be understood. In the same way programming tools vary in usability and fitness for various tasks, these sketched notations may have combinations of affordances and trade-offs. Building datasets of UI sketches and systems that use sketched notations, without first understanding how best to create the sketched notations themselves, risks the future usability of these systems.

To better understand the affordances and usability implications of sketched notations for UI design and development, and to inform the design of future human-centered tools that use these notations, we use Green's Cognitive Dimensions of Notations framework [5] to analyze two potential sketch notations for UI design. For the first notation, we heuristically evaluate an existing dataset of UI sketches where the data producers were given a fixed set of sketching rules to follow, for example, "draw image elements as boxes with X's in them". For the second notation, we embrace the fact that part of the benefit of sketching is that it allows people to represent whatever they want, however they want. The notational implications of maintaining this aspect of sketch for multimodal UI design tools are unstudied. In order to analyze intuitive sketches as a notation, we conduct a study of 21 individuals sketching UIs without any constraints on how sketched elements should look. The data from this study forms our second notation, which we examine to understand the feasibility and trade-offs of enabling a less constrained approach to UI sketching.

This paper makes the following contributions:
1) An open dataset containing a range of intuitive sketched notations produced from a study of 21 individuals intuitively sketching UIs with minimal guidance.
2) A cognitive dimensions analysis of two sketch notations showing that while the open-ended approach to sketched notations lacks the consistency of the closed approach, the open-ended approach is more usable and expressive.
3) Recommendations for the development of systems that can interpret the rich relational and contextual information embedded in intuitive, flexible sketched notations.

## II. RELATED WORK

This paper draws upon prior work on sketch in UI design, both for ideation and system specification.

### A. Sketch-Based UI Design Systems

Numerous prior works have identified the important role of sketch in design processes, particularly in UI design [1], [3], [6]. The main advantage of sketch is that it allows designers to rapidly express their ideas without investing too much effort, enabling them to consider a variety of designs without overthinking specific design or implementation details [7].

HCI researchers have long been fascinated with the potential of sketch as a way to define UIs. Traditionally, UI design begins with visual prototypes that are useful to designers but do not generate or integrate with code that defines system behavior [2], and often do not represent system behavior at all [8]. This visual-behavioral disconnect is not ubiquitous, and prior HCI work such as DENIM [1] addresses this limitation by allowing users to map out the flow of their interfaces through interactable sketches, while SILK [9] is able to interpret a set of predefined visuals as UI elements and transform them into a code skeleton. Similarly, JavaSketchIt introduces a grammar that defines a set of sketches and gestures that are recognized and turned into UI elements [10]. These projects predate recent advances in computer vision and other ML methods that could, in theory, support a greater diversity of ways of designing applications and their interfaces. Newer approaches (e.g. Doodle2App, Sketch2MultipleGUIs) use heuristics or deep learning approaches to interpret a predefined set of sketch notations to generate UIs [11], [12]. Each of these prior works provide (i.e. pre-define) a notation for how users should sketch certain elements.

### B. Sketched UI Element Datasets

While most research on sketch-based UI design tools predates the widespread adoption of ML, recent work recognizes the potential of ML to support the construction of more powerful sketch-based design tools, developing datasets for training ML models to recognize sketch-based UI elements [13]–[15]. The process of each dataset's development has major implications for what it contains, and therefore the notation it prescribes for a future system. While each dataset is different, they share certain commonalities. The ImageCLEF drawnUI dataset and the Swire dataset were both created by data producers following specific instructions provided by researchers for how to draw each UI element [13], [15]. The UISketch dataset consists of sketches drawn without notational instructions, but only includes isolated UI elements (out of the context of a larger design) [14] unlike the ImageCLEF drawnUI and Swire datasets which contain whole UI sketches. Each of these datasets prescribes a slightly different possible notation. For the purposes of our analysis, we select one example dataset, the ImageCLEF drawnUI dataset, to represent one possible type of notation of UI sketches. This dataset contains around 4000 labeled UI wireframe sketches for researchers to use to train interface sketch recognition models [15]. We chose this dataset because of the availability of the representational constraints given to data producers, which are viewable on the project website. Figure 1 shows the instructional examples given to the ImageCLEF data producers. They were instructed to draw elements in specific ways, and were shown "incorrect" examples of various UI element sketches to avoid. The set of instructions given to data producers therefore encompasses the total legal notation.

Fig. 1. The ImageCLEF drawnUI dataset instructions for people drawing the wireframes. Data producers were given 'Correct' and 'Incorrect' depictions of each UI element. These sketches together encompass the whole notation.

### C. Cognitive Dimensions

The cognitive dimensions framework is a way to broadly evaluate the usability of a system. The 14 dimensions presented by Green et al. relate to various aspects of cognitive effort that may be required from different approaches to system design [5]. Table I shows the dimensions and their definitions. Each cognitive dimension is not necessarily a metric to be maximized, as improving one dimension may come at the expense of others. Instead, analysis through the dimensions' lenses allows for a greater understanding of a system, the ways it may be challenging to use, and opportunities for improvement. Green emphasizes that systems are combinations of notations and environments, and each can and should be designed to support the other towards the system's larger goals [16]. Analysis of the cognitive dimensions of a notation also provides insight into how the notation's environment should be designed. For the purposes of our analysis, the environments are the tools that interpret sketches to generate UIs, and the notations are the sketches interpretable by those tools.

| Dimension | Definition |
| --- | --- |
| Abstraction Gradient | Types and availability of abstraction mechanisms |
| Closeness of Mapping | Closeness of representation to domain |
| Consistency | Similar semantics are expressed in similar syntactic forms |
| Diffuseness | Verbosity of language |
| Error-proneness | Notation invites mistakes |
| Hard Mental Operations | High demand on cognitive resources |
| Hidden Dependencies | Important links between entities are not visible |
| Premature Commitment | Constraints on the order of doing things |
| Progressive Evaluation | Work-to-date can be checked at any time |
| Provisionality | Degree of commitment to actions or marks |
| Role-expressiveness | The purpose of a component is readily inferred |
| Secondary Notation | Extra information in means other than formal syntax |
| Viscosity | Resistance to change |
| Visibility | Ability to view components easily |

TABLE I
THE 14 COGNITIVE DIMENSIONS [5]

### III. COGNITIVE DIMENSIONS OF THE IMAGECLEF DRAWNUI DATASET NOTATION

#### A. Description of the System

In order to discuss the cognitive dimensions of a sketch-based UI design system that utilizes the ImageCLEF drawnUI dataset rules as its notation, we must first describe the imagined use and behavior of the system. This hypothetical system, which we refer to as the FixedSketch system, uses the drawnUI rules as its notation and provides a drawing environment that uses a ML model trained on the drawnUI dataset to recognize users' sketches. We stipulate that users can draw any of the 21 defined UI elements as depicted in the "correct" column of Figure 1 in a digital drawing environment. As they draw, their sketches are interpreted by a vision model and used to build a functional UI in near real-time with elements in the same relative locations as in the sketch. The environment has all the features of a typical digital drawing tool, including the ability to draw, erase, move, and resize elements on a canvas of adjustable size.

#### B. The Cognitive Dimensions

We now consider the various cognitive dimensions of the hypothetical FixedSketch system.
**Visibility** Visibility describes the relative ease of viewing the components of a design. The FixedSketch system is highly visible due to the nature of sketch itself. A drawing canvas is a high visibility environment that allows the user to see their whole design at once. As a result, cognitive effort does not have to be spent on mentally keeping track of elements.
**Premature Commitment** Premature commitment concerns the extent to which decisions have to be made in a specific order, and cannot be easily undone or re-ordered. Sketching, especially in a digital context as with FixedSketch, has low premature commitment; Users can adjust and rework their designs, and create elements in many orderings. This reduces the up-front need to anticipate potential conflicting decisions, and instead allows users to continuously adjust their designs.
**Abstraction Gradient** Abstraction gradient describes the potential means of abstracting the notation available to the user. The FixedSketch system is what Green and Blackwell would call "Abstraction-Hating", meaning the system does not allow users to define new abstractions [5]. The system's ability to interpret sketches hinges upon its ability to recognize them, and new representations of elements or combinations of elements will not be supported by the vision model, which can only recognize the 21 elements when they are drawn as specified by the drawnUI rules. While novice designers may prefer to work within a limited set of concrete options [5], for more advanced designers, abstractions serve as a way to both make their workflow more efficient by collecting elements together in larger, easily repeatable patterns, and allow them to represent more complex designs in more simplified, manageable ways.
**Progressive Evaluation** Progressive Evaluation refers to the ease with which the user can examine their progress. FixedSketch supports progressive evaluation, since in order to check their progress users need only look at their sketch and the generated UI, which is updated progressively as they sketch. This allows users to verify that the outcomes of their choices are as they expect, and make adjustments when necessary.
**Viscosity** Viscosity concerns how change-resistant a system is. For the FixedSketch system, the benefits of high visibility and low premature commitment come with the caveat that the user must be satisfied with the 21 defined UI elements. While the act of sketching on a canvas is low viscosity, the system itself is highly viscous because users cannot introduce new elements or modify existing ones. This limits the usability of the system for users who want to express more diverse designs.
**Consistency** Consistency refers to the degree to which similar elements are represented similarly. A consequence of the FixedSketch system's abstraction-hating approach is that this system is extremely consistent. There are only one or two ways to draw any given UI element, and as such there will not be great variability between sketches as to how each element appears. The set of 21 element sketches themselves are also relatively consistent. For example, text is represented as a squiggly line while a link is represented as a squiggly line in brackets, which maintains a consistent representation of text across the multiple elements that contain text. This consistency can make it easier for multiple people to interpret each others' sketches and contribute to a single design.
**Provisionality** Provisionality describes the ease with which actions can be undone. As with premature commitment where elements can be created in any order, FixedSketch is also highly provisional—it is relatively simple to erase, reposition, or otherwise adjust an already sketched element in a digital drawing environment. This reduces the cognitive effort required to plan ahead, as prior decisions can be easily changed.
**Role-expressiveness** Role-expressiveness, or the ease with which one can infer the purpose of a component, is variable with the FixedSketch system. Text elements are depicted as the word 'text' or squiggles which are relatively recognizable as being text, however one cannot write the actual text content, such as "Home", envisioned for their design. Doing so is supported as a form of progressive evaluation in prior work like DENIM [1] and is more role-expressive. For image elements, the "correct" depictions show boxes or circles with X's in them, which aren't obviously recognizable as images. The "incorrect" examples of images are skeuomorphic, showing a sketch of a mountain scene and a person which are more reminiscent of how an image might look. Whether or not someone recognizes any of these sketches as a given element is somewhat subjective, and may vary from person to person. Consequently, additional learning effort is required from users in order for them to adopt the notation rules.
**Error-proneness** Error-proneness concerns whether the notation invites mistakes. The existence of the "incorrect" column of the FixedSketch notation document in Figure 1 suggests that the notation system designers recognize that there is a larger range of intuitive representations than the system allows. This opens users up to mistakes, where they may draw an element wrong and be misunderstood by the system.
**Hard Mental Operations** Hard mental operations are tasks that require high cognitive effort. In the case of the FixedSketch system, users have to disregard their own intuitions for element representations and instead remember a fixed set of representations. Especially during the novice phase, memorizing, recalling, and correctly drawing this arguably arbitrary set of representations with variable role-expressiveness requires significant cognitive effort, and invites error.
**Secondary Notation** Because this set of sketches encompasses the whole notation, this system does not support any additional way to convey meaning, or secondary notation. Secondary notations that are not part of the main notation of a system are useful because they support comprehension. A secondary notation for UI sketching would allow a user to provide additional information describing the design and purpose of the interface, making it easier for them to remember their progress or show their design to others. Instead, with this notation, all sketches are intended as UI elements, and will be interpreted as such by the system.
**Closeness of Mapping** A close mapping system is one that reflects the relevant domain. In some ways, the FixedSketch system does map well to the domain of UI design. A sketch made in this system would look similar to a low-fidelity UI mock-up, and drawing a UI element in the system maps to the creation of a UI element. However, it does not fully map to UI design, as it does not allow for representation of custom UI *behavior*, which is a key aspect of the UI design domain.
**Diffuseness** Diffuseness describes the degree to which the notation uses excess descriptiveness, or verbosity. While in language verbosity refers to using excess words to describe something, in this context verbosity can describe using excess sketched components. For the FixedSketch notation, and arguably any sketch notation, its diffuseness is tied to its role-expressiveness. Many of the elements in the notation are fairly detailed in order to disambiguate element types. A set of 21 symbols composed of one or two lines each would be less verbose, but would lose all role-expressivity and require hard mental operations to remember and repeat. Making each element as specific as possible in order for each element's role to be obviously apparent would involve a more verbose representation that is more time consuming for the user to depict, and a very detailed notation may incur additional error-proneness. In contrast, a more verbose representation in sketch may actually be easier to interpret, unlike with written language. Whether any of the representations used in the FixedSketch system are too detailed, or not detailed enough, is somewhat subjective.
**Hidden Dependencies** We did not directly observe any hidden dependencies, or invisible links between entities, within the notation itself. There are many ways hidden dependencies could be present in the FixedSketch system, for example the invisible link between a sketch and a model's classification of that sketch, but that dependency, and the extent to which it is hidden, would be the result of design choices for the system's environment rather than the notation.

*C. Summary*

With the hypothetical FixedSketch system, many of the affordances of sketch are curtailed by the structure of the notation, whose rigidity can make an otherwise expressive and flexible design process error-prone and scoped to a limited representational space.

## IV. STUDY OVERVIEW

We wish to develop systems that are flexible enough to work for people whose intuitions about how to visually express their ideas differ from one another's (and from our own). We want to minimize our imposition of representational constraints on peoples' design processes, and aim to create tools that support top-down, bottom-up, or more exploratory, bricolage [17] approaches. Progress requires us to understand how different people intuitively sketch UI elements and UI behaviors, and develop insights about intuitive (sometimes called "natural" [18]) notational practices to inform new approaches to sketch-based design tools. Accordingly, in this study, we investigate the following research questions:
**RQ1** – What representations of UIs do people intuitively sketch?
**RQ2** – What are the cognitive dimensions of a notation defined by these representations?
**RQ3** – What are the implications of these representations for the design of sketch-based UI design tools?

In order to answer these questions, we conduct a study of 21 individuals sketching UIs intuitively—without any notational instructions. We treat the set of our participants sketches as a notation, and imagine a hypothetical FlexiSketch system that supports this notation in a similar drawing environment as the FixedSketch system, but also allows user correction of system misinterpretations of their sketches. We then consider the cognitive dimensions of this system and the implications for how future, more flexible design tools should be designed.

## V. METHODS

### A. Recruitment

Our study protocol was reviewed and approved by our university's Institutional Review Board. We solicited participation by individuals in our university community with experience drawing, in UI Design, and/or programming in order to capture a broad cross section of the population. We advertised our study in art, architecture, and computer science spaces in order to find participants with varied prior experiences. We obtained consent for participation from every participant.

### B. Procedure

In our 3-phase study, we asked participants to draw four interfaces in the iOS app Freeform on an iPad, answer follow up questions, and label their drawings.

*Phase 1) Drawing:* The first three drawing tasks involved participants sketching existing interfaces, giving them the opportunity to become familiar with the idea of sketching UI elements and develop their own personal shorthands and representations. This also allowed for direct comparison between our participants' sketches of the exact same designs. The three UIs that participants copied increased in complexity, with Task 1 being the Google homepage, Task 2 a personal website, and Task 3 a library website (Figure 2). These sites were chosen because they were likely to be familiar to participants, and also collectively contained a variety of UI elements, such as buttons, images, text, etc. We conjectured that this familiarity was advantageous because it reduced the risk that participants would need to devote effort to understanding the provided UIs, or misunderstand them, which would create validity risks for our later data analysis. In Task 4, the final drawing task, we asked participants to draw an interface of their own design. The purpose of this task was to see how participants drew UIs from their imaginations, unconstrained by existing interfaces. We avoided instructing participants on any workflow to use or content to include, in order to capture the intuitive behaviors of novices sketching and designing UIs.

*Phase 2) Follow Up Questions:* After participants completed the drawing tasks, we asked them how they felt about sketching as a way of prototyping an interface. We also asked what suggestions they had for features or affordances of a future design tool that would incorporate and automatically interpret sketches, in order to capture what support might be necessary for this type of sketching in future tools.

*Phase 3) Labeling:* Finally, we went through each of their drawings with them and asked them to describe what each sketched UI element was. All participants were given a list of UI element names and visuals to support this labeling task. This list was only provided after participants completed the initial drawing task so as to not influence their chosen representations. Participants were given freedom to interact with the actual website to confirm their label decisions. When participants struggled to identify the names of certain elements, we suggested possible labels.

### C. Analysis

With the labels provided by the participants, we cropped and annotated the sketches element-by-element. We then quantitatively analyzed the relative occurrences of certain labels and representations, as well as qualitatively characterized the visual qualities of the representations and participants' responses to the study questions. To analyze participants' verbal answers about their experiences and preferences, we analyzed transcripts of participants' responses using affinity diagramming to group them into common themes.

## VI. RESULTS

### A. Participant Characteristics

Participants joined the study with varying degrees and combinations of prior experience with programming, drawing, and user interface design. The majority (86%) of participants reported recent drawing experience. Two thirds (67%) of participants had interface design experience and around 71% of participants had programming experience. Just over half of participants (57%) had both interface design and programming experience, and indicated previously having made a website, game, or app. We did not identify any significant differences in drawing behavior based on prior experience.

### B. What Participants Drew

RQ1 asks *What representations of UIs do people intuitively sketch?* Our 21 participants produced 84 UI sketches with 4105 total UI elements. Each element appeared in context with other elements and was labeled in detail, enabling in-depth analysis on both appearance and use in the larger design.

*1) Element Appearance:* To investigate RQ1 we broke down each type of UI element (as labeled by participants) into categories based on the element's sketched appearance.

*a) Text:* We identified six unique ways participants sketched Text: *Literals* (i.e a word written out), *Bounding boxes, Lines, Squiggles, Dots,* and *Xs*. These six representations were often combined with one another within a single Text element. Figure 3 shows a sample of the different Text representations and their combinations drawn by participants.

*b) Images:* Similarly, we identified five different categories of Image sketches: *Detailed* (they drew a high-fidelity copy of the source image), *Notional* (they drew a symbolic, low-fidelity representation of the image content), *Placeholder Symbol* (they drew a symbol to represent the existence of an image, but not relating to the image's content), *Placeholder Text* (they wrote a word to designate the item an image), and *Placeholder - None* (they drew only a bounding shape). Figure 4 shows a selection of different Images drawn by participants.

There was significant range in usage of each of these Image representations by participants. All participants, for example, drew the profile image in Task 2, the personal homepage.

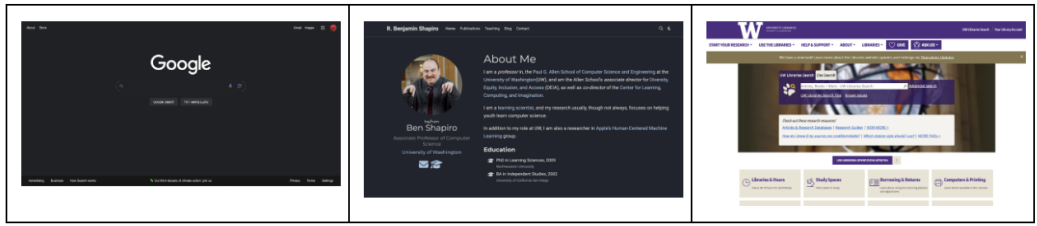

Fig. 2. Interfaces of increasing complexity that participants sketched during the study.

Fig. 3. The different ways participants represented Text including Literals, Bounding Boxes, Lines, Squiggles, Dots, Xs, and combinations of multiple representations. The fraction of total text representations drawn with each type is noted next to the representation name.

Fig. 4. The different ways participants represented Images, including as Detailed Images, Notional Images, Placeholder Images with Symbols, Text, or just Bounding Boxes. The percentage of total images represented with each method is noted next to each type.

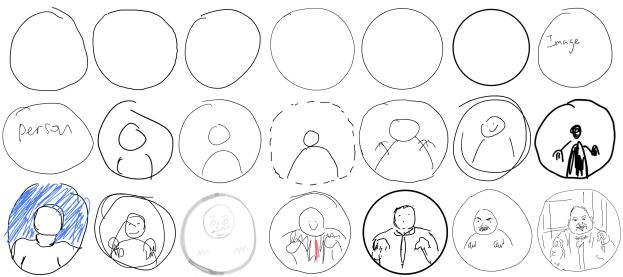

Fig. 5. Depictions of the circular profile photo on the copied personal homepage. Representations varied from a simple placeholder circle to a fully detailed portrait.

Fig. 6. The different ways people represented Icons. Most Icons were drawn as the specific Icon the participant wanted, though some were represented with placeholders.

Fig. 7. (a) Example ways people represented Cards, Line Breaks, and Buttons. (b) Example sketches of Dropdowns, Lists, Sliders, and Text Inputs.

Figure 5 shows each unique way this Image was represented among the twenty one participants, ranging from an empty placeholder to a detailed portrait. While some participant sketches are more reminiscent of wireframes, others' UI sketches are more akin to artistic illustrations.

*c) Icons:* Icons were drawn two ways: either the participant sketched their desired icon in detail as it was meant to appear (82.8%), or they drew a placeholder to denote an icon element (17.2%), mirroring the detailed and placeholder approach to images. Figure 6 shows example icon sketches.

*d) Containers:* We used the label of 'Container' to describe elements that conveyed specific bounding constraints. This included Cards, which are visible bounded regions within a UI, Line Breaks, which delineate zones within a UI, and Buttons, which are interactive elements that are specifically bound in a visible container. To distinguish whether an element was a Button as opposed to a link or some other interactive element, we defined a Button as an interactive element that was bound in an explicit container, that is, a container that was meant to be visible in the final UI. Often, participants drew an unbounded Text, Image, or, Icon element, but described it as a button during labeling. We labeled these elements as being implicitly interactive, meaning there was nothing about the way they were sketched that distinguished them as interactive. Figure 7 (a) shows examples of sketched Container elements.

*e) Other Elements:* There were fewer instances of the remaining four elements, Dropdowns, Text Inputs, Lists, and Sliders, and less variation in their depictions. Figure 7 (b) shows examples of sketches of these elements.

## C. Cognitive Dimensions of FlexiSketch

These representations of UI elements (the notation) and their use in context (the environment) could be the basis of a system, which we refer to as FlexiSketch, with a unique set of cognitive dimensions. To account for the open-ended nature of the notation, this hypothetical system would have the same drawing environment as the FixedSketch system with the addition that users can correct the system's interpretation of their sketch if misunderstood. To answer RQ2—*What are the cognitive dimensions of a notation that consists of these representations?*—we analyze this notation using the cognitive dimensions framework.

**Visibility** Like the FixedSketch system, the FlexiSketch system is high visibility, meaning components of the design are easily viewable.

**Premature Commitment** The FlexiSketch system's lack of premature commitment, or degree to which actions have to be taken in specific orders, was apparent in the two distinct approaches to sketching we observed. Some participants primarily sketched shapes as representations of elements (not varying stroke or color), while others preferred to depict the visual appearance of the interface they were copying or imagining (matching stroke and color features of UI elements). The cognitive effort alleviated by the visibility of the system is different for each approach—the user of the first approach wants to see their design's structure, and the user of the second wants to see their design's appearance. Enabling either approach reduces the feeling of premature commitment for users with different preferences.

**Abstraction Gradient** The abstraction gradient, or the types and availability of abstraction mechanisms, was broad and expanding for the FlexiSketch notation, as participants invented a large range of abstractions while sketching. As the copied designs became more complex, and as participants became more familiar with the drawing task, use of abstracted representations for text increased. Instead of writing out the text literally, they used mixtures of shorthands including squiggles, lines, dots, bounding boxes, and Xs, as shown in Figure 3. This increase occurred both in the total amount of text elements represented with shorthand, and in the number of unique abstractions used. The average number of distinct text representations used in the first three tasks were 1.3, 2.5, and 3.4, respectively. When participants used literals in their original designs, it was often to represent important conceptual details, such as headers and the names of navigation bar elements, defining the overall content of the page literally, then leaving smaller descriptive elements in shorthand.

**Progressive Evaluation** This abstraction behavior suggests that the choice of literal versus shorthand text in the design process served to support progressive evaluation, or the ease of checking one's progress. Participants retained detail for certain elements, allowing them to quickly understand and evaluate the most important aspects of their design, while abstracting away less important elements. This effect can be seen in Figures 3 and 4, where more and less detailed versions of image and text elements are shown.

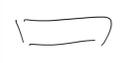

Fig. 8. Examples of Text, Image, Icon, Card, and Button elements represented in very similar ways, making them difficult to distinguish.

**Viscosity** Inherent in the setup of this FlexiSketch notation is its low viscosity, or low change-resistance. Within a single individual, we observed the use of many different representations for the same types of elements, and the set of chosen representations changed and grew for each individual. This variety is apparent in Figures 3, 4, 5, and 6.

**Consistency** The eager use of abstraction by our participants introduces trade-offs for consistency, or the use of similar representations for semantically similar elements. Participants did not prioritize uniqueness between representations, often representing different elements using the same sketch, even within a single drawing. Figure 8 shows some examples of these overlapping representations. This phenomenon exposes the shortcomings of only using element-based sketch understanding models, as these representations are visually indistinguishable. It also indicates that notational requirements for one-to-one mappings between a sketch and UI element does not reflect sketching in practice.

**Provisionality** While provisionality, or the ease of undoing an action, is already high for sketch, the flexible nature of the FlexiSketch notation would also increase the provisionality of the system, as users would not need to commit to any one notational choice for the whole design process.

**Role Expressiveness** While navigating the inconsistencies of the FlexiSketch notation may appear to require significant effort from the system, the ability to draw whatever they wanted came with beneficial trade-offs for participants in terms of role expressiveness, or the ease of determining the purpose of a component. The degree of role expressiveness of their sketches came down to the design choices of each participant, with some prioritizing detail more than others, as shown in Figures 4 and 5. Regardless, participants retained the ability to be as role expressive as they saw fit, representing elements in ways that were meaningful and interpretable to them.

**Error-Proneness** While the notational flexibility of FlexiSketch gave participants the opportunity to represent elements however they wanted, removing the threat of error from sketching elements "wrong", we identified that participants rarely positioned elements in their exact positions

Interface 3 navbar as it appeared to participants:

Interface 3 navbar as sketched by participants:

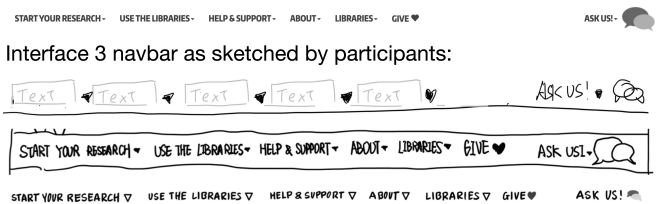

Fig. 9. The navigation bar on interface 3 and a sample of sketches of the same navigation bar drawn by participants.

relative to the copied sites using perfect padding and spacing, potentially inviting error. Figure 9 shows an example of this effect using the navigation bar on the third interface. These participants represented most if not all of the correct elements on the navbar. However, they did not leave the correct amount of space between the 'Give' element in the left group of items and the 'Ask Us' element in the right group of items. This indicates that in their original designs, they may not be able to depict layout and spacing as precisely as they intend.

**Hard Mental Operations** By not requiring the memorization and recall of any element representations, the FlexiSketch approach reduces the hard mental operations, or demand on cognitive resources, introduced by sketching rules. The unconstrained approach reduced cognitive effort by allowing the immediate expression of ideas, instead of having to interrupt creative flow in order to find the correct way to communicate.

**Secondary Notations** Secondary notations, or means of representing information outside of formal syntax, is difficult to define for the FlexiSketch notation since the notation has no formal syntax. However, participants drew sketched representations that went beyond UI elements, which we consider a form of secondary notation for this context. Though they were instructed to sketch the elements on the given interfaces, some participants sketched additional details to communicate their intentions—sketches that were not themselves UI elements.

The most common occurrence of additional sketches came in the form of annotations, typically where participants would self-label their elements with the name of the type of element. Other forms of annotations showed edits, such as a caret pointing where text should be moved or inserted, instead of actually performing the move or redrawing the sketch. Another type of communicative sketch was dashed, colored line breaks used to show invisible padding boundaries. In the fourth sketch task, after participants sketched their original design, they were instructed to add a page to their designed interface and show in sketch how the two pages connected. Participants represented these interactions by sketching in five unique ways: *Arrows* (an arrow pointing from a button to a new page), *Name Correspondence*, *Icons* (matching Icons drawn in two locations to show a connection), *Colors* (a Text element and its corresponding linked page circled in a shared color), and *Annotations* (the word "click" placed next to an Arrow between an element and a page).

The final type of sketched communications were used to convey commands. For example, a participant trying to depict a group of icons drew a single placeholder icon, then wrote 'x6' next to it to indicate it should be copied six times. The 'x6' sketch was not itself a UI element, but an expression of a command, in this case copy/paste, that they would want FlexiSketch to interpret and execute. Other expressions of this 'copy the previous element' appeared in multiple forms, including labeling an element with a number and then enumerating number values where they wanted the element repeated. Other participants drew an element and then wrote '...' to indicate a repeat of the drawn element. When one participant drew multiple pages with a common navigation bar, they drew the first navigation bar in detail then drew a squiggle in the second navigation bar where they wanted the elements to be the same. This indicates that it is intuitive to utilize secondary notations that do not depict UI elements within the UI sketch. Unlike pen gestures, which are drawn as strokes but simply trigger a specified action, these types of sketched commands form part of the sketch itself as a secondary notation.

**Closeness of Mapping** Secondary notations can also improve the system's closeness of mapping, or how well it reflects the UI design domain, as they would allow users to fill in functional details of a UI, making the sketching task map fully to the the task of UI design and implementation.

**Diffuseness** As with the FixedSketch system, the diffuseness, or verbosity of language, of a sketched notation is somewhat subjective. The FlexiSketch notation is theoretically infinitely diffuse—users can draw anything, including increasingly complicated and detailed depictions of UI elements. In language, verbosity describes when more words are used than necessary. The concept of "necessary" in the domain of sketch is debatable; a detailed sketch conveys more information than a simple one, and due to the visibility of sketch, a more verbose notation does not necessarily translate to more cognitive effort. While most participants represented elements that were akin to elements from a prototyping tool, there were some truly unique designs. Figure 10 shows one such design created by a participant in their original drawing task. They depicted an iceberg surrounded by animals and a submarine. They explained that the iceberg should melt when you click it, and that the animals on icebergs were buttons that are floating on an animated water background. While this design may be difficult to represent using a more terse notation, it is not a difficult design to implement.

**Hidden Dependencies** Hidden dependencies describes notations where entities are linked in ways that are not readily apparent. As with many of the cognitive dimensions of the FlexiSketch system, the degree to which a dimension applies depends on the choices made by the participant while sketching. We observed instances where some participants sketched in such a way as to introduce hidden dependencies in their designs when sketching groups of similar elements. For example, these participants would sketch the first element in detail, then sketch generic placeholders for the remaining elements. Viewed together, these placeholders reference the first element in the series, but examining each element alone obscures the existence of a reference to the prior element.

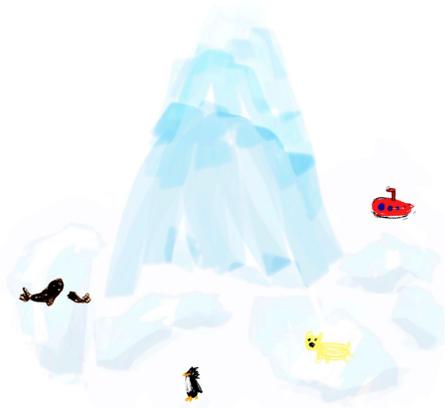

Fig. 10. Participant 19's original UI design, depicting an iceberg that should melt when clicked surrounded by buttons depicted as animals and a submarine.

*D. Summary*

The FlexiSketch approach enabled participants to sketch according to their intuitions. The resultant notation reflects a human design process—it is at some times inconsistent and imprecise, while being highly expressive.

## VII. Discussion

We have analyzed two notations, and two (otherwise-identical) hypothetical systems built using them, to deepen our understanding of the usability and expressivity of notational alternatives for sketch-based UI design and development tools. FixedSketch and FlexiSketch are proxies for two design system extremes—either the user can draw one fixed set of symbols (closed-ended) or they can draw anything at all (open-ended). Our analysis of the first extreme draws on a notation from prior research, while our own data support our analysis of the latter extreme. To answer RQ3—*What are the implications of these representations for the design of sketch-based UI design tools?*—we summarize our findings about the affordances of each notation based on their cognitive dimensions, provide recommendations for future sketch-based design tools grounded in our analysis, and identify needs for further research.

*A. Summary of Findings*

Both notations benefit from the baseline visibility, provisionality, progressive evaluation, and low premature commitment of sketching. The main benefit of the closed approach is consistency. On the user side, this means that sketches made by different users will use the same set of representations, potentially facilitating collaboration. On the system side, this consistency makes it feasible to use *element-based* sketch recognition models, trained to differentiate between a fixed set of element representations, to build sketch-based systems.

The sketches produced by our study participants are vastly different and more varied than any closed notation could capture. This indicates that closed approaches, and existing datasets that embody them, do not satisfactorily reflect intuitive sketching behaviors. If sketch-based UI design tools are built with these datasets alone, they are likely to poorly support users' intuitive practices. Using sketch as a modality for UI design is not simply about representing elements as strokes on a canvas; sketch is a particularly expressive medium. While tools could provide users with closed notations and require users to learn them, doing so would prevent users from taking advantage of the main benefit of sketch: its felicitous embrace of people depicting whatever they want, however they want.

The open-ended approach has many usability affordances which reflect the qualities of sketching that are appealing to designers. Users do not have to memorize, recall, and accurately depict a set of representations, reducing hard mental operations and error-proneness. Users can depict the exact design they are imagining, including specific aesthetic details, enabling role expressiveness and supporting progressive evaluation beyond the baseline closed approach. An infinite abstraction gradient and low viscosity allows users to invent and transform notations to suit their changing needs and design vision. The potential for secondary notations allows users to convey significantly more information in their designs. The notational inconsistencies of this approach are only an issue if systems are not designed to handle them. Supporting open-ended approaches to UI sketching requires the development of new interaction techniques and models that enable ambiguous and ever-changing inputs.

*B. Recommendations for Future Sketch Systems*

A key advantage of closed-ended notations is that they can be designed to ease the implementation of element-or stroke-based *recognition* models, but, as discussed above, this comes at the cost of demanding greater user work (e.g. memorization and recall) and diminishing expressivity. Ultimately, that is a machine-centered approach, rather than a human-centered one. A human-centered approach necessitates developing new *sketch understanding* methods that are capable of handling these heterogeneous and ever-evolving, open-ended notations. We draw on our cognitive dimensions analysis to present the following recommendations for how to design such systems.

**Recommendation 1: Support relational notations using context.** Our data analysis revealed common patterns for how individual elements are embedded in, and gain meaning through, their juxtaposition with other elements. For example, once participants developed shorthands to represent Text, they continued to mix Literals and Shorthand representations for Text within a single UI drawing instead of replacing Text with Shorthand entirely. When they did so, Text represented with Shorthand was more often used to represent paragraph or label Text, and headers often remained as literals. Patterns also appeared for repeated elements, where participants often drew the first element of a series in detail, and then used placeholders for the remaining elements. Taking these patterns into consideration, we propose a conceptual shift in what a sketched notation for a UI is, and how we should build systems to parse those notations. Instead of building systems that synthesize interfaces piecewise based upon what each element looks like, sketches must be modeled as relationally-rich inscriptions. Much like UI understanding [19], sketch understanding systems should take advantage of spatio-temporal contextual information to interpret open-ended notations.

**Recommendation 2: Support notational change at the individual level.** The notational variability we observed cannot simply be accounted for by learning a fixed notation per individual, as our participants regularly changed their notations over time. Sketch-based design tools that support open-ended notations must incorporate online learning to account for constant notational change. Systems must be able to learn variations in how to represent already-defined UI elements, as well as support user definition of new kinds of elements or multi-element abstractions. When people can express their intuitions before having to think about notational constraints, they are able to create non-standard elements, pushing the frontiers of UI design. It is conceivable that designs like the iceberg shown in Figure 10 are rare in part due to the prescriptive nature of prototyping tools and sketching notations. Moving from (fixed) element-recognition methods to dynamic sketch understanding and software synthesis techniques is a prime opportunity to apply generative AI methods to augment human capabilities. Prior work has identified that generative AI can have a homogenizing effect, where outputs converge to common baselines [20]. We can reduce this risk by providing mechanisms for users to customize sketch understanding models, so that rather than changing their sketches to comport with what models already understand (amplifying homogenization), people have agency to make models understand them.

**Recommendation 3: Support novelty through user corrections.** Systems can also embrace variations in how people represent their ideas in sketch by providing users with avenues for correction when systems misinterpret their sketches. By doing so, they provide training data to improve the system's sketch understanding. Representation and incorporation of corrections likely will need to happen at two levels: First, a user-provided correction to a sketch segment's meaning must be treated as fixed and never be reclassified. Second, systems should learn from these corrections to adapt to each user's evolving notation. Recent advances in the ability to fine-tune generative models with reinforcement learning from human feedback (RLHF) offer the possibility to train models to adjust based on user corrections [21], including recent prior work introducing novel techniques that can personalize to diverse user preferences [22].

**Recommendation 4: Support layered notations.** Future systems should support layered (secondary, tertiary, ...) notations that allow users to intuitively describe envisioned systems' behaviors, and even expected patterns of user behavior (e.g. by allowing users to sketch different routes between screens and automatically synthesize storyboards depicting different scenarios). Systems that could recognize these additional notations, and support users in associating meaningful operations with them, would enable a new paradigm of intermixed visual and logical sketched expression where appearance, context, use cases, and the flow of time are all taken into account when sketch systems interpret design intent.

## VIII. FUTURE WORK

A challenge for open-ended sketched notations is that because they are so flexible, each person may sketch in such a way as to introduce different cognitive dimensions to their individual notation, as shown by our analysis of hidden dependencies, role expressiveness, and overall inconsistency. We should investigate if there are optimal notational middle grounds to help users develop notations between the extreme open- and closed-ended notations of this study to balance expressive power, ease of use, productivity, and computational understandability. Examples of others' notational choices may help users to find ways of expressing their ideas that feel right and allow them to be more easily understood by future design tools. Such examples could be curated from a library of others' designs or drawn from sketch understanding training examples that are far apart in the model's learned embedding space.

We do not think future design tools should be limited to visual representations alone, and must support dynamic design processes that intermix drawing with speech and gesture. Supporting open-ended, multi-modal design processes will require the invention of new interaction techniques and models that enable adaptation in response to ever changing inputs and input channels; what someone expresses with a drawn arrow in one moment might be expressed in the next by pointing or tapping and saying "from here to here." To achieve this vision, researchers will need to devise methods to combine context-sensitive sequence modeling methods (e.g. transformers) – to glean meaning from users' multi-modal inputs within spatio-temporal context – with reinforcement learning methods – to incorporate user corrections when models mispredict users' intentions. New modeling methods must be combined with HCI research that investigates how these models' interpretations of users' inputs are presented to them, how people make use of different interfaces for correcting mispredictions, and more.

Every modality a system supports changes the available notation in some form, and we hope that future work will study people's intuitive expressions with different combinations of modalities in order to develop models and AI approaches for incorporating modal flexibility into software design and development tools. Our small dataset of participant sketches, combined with our analysis, represents the first step in this process, providing insights and initial training data for newer, more flexible sketch-based systems. The dataset is available at https://huggingface.co/datasets/samhross/flexiSketch. We invite others to use and add to it, to increase its generalizability.

## IX. CONCLUSION

This paper presents an analysis of the cognitive dimensions of two different approaches to sketched notations for sketch-based UI design systems. We compare an existing dataset's notational rules to a study of individuals sketching UI elements without constraints, in order to make recommendations for future systems that support intuitive creative expression. We hope to contribute to a future of multi-modal design tools that support both advanced software development as well as unconstrained creative expression.

ACKNOWLEDGMENT

This material is based upon work supported by the NSF Graduate Research Fellowship under Grant No. DGE-2140004. Any opinions, findings, and conclusions or recommendations expressed in this material are those of the authors and do not necessarily reflect the views of the National Science Foundation.